\documentclass[aps,prl,twocolumn,showpacs,superscriptaddress,groupedaddress]{revtex4-1}

\newcommand{\ket}[1]{\left|#1\right\rangle}

\newcommand{\Ca}{Ca$^{+}$}
\newcommand{\Ba}{Ba$^{+}$}

\newcommand{\Yb}{Yb$^{+}$}
\newcommand{\Sr}{Sr$^{+}$}

\newcommand{\Mg}{Mg$^{+}$}

\usepackage[utf8x]{inputenc}
\usepackage{graphicx}
\usepackage{dcolumn}
\usepackage{bm}
\usepackage{amsmath}
\usepackage{amssymb}
\usepackage{subfigure}
\usepackage{subfloat}

\begin{document}

\title{Direct spectroscopy of the $^2$S$_{1/2}-^2$P$_{1/2}$ and $^2$D$_{3/2}-^2$P$_{1/2}$ transitions and observation of micromotion modulated spectra in trapped \Ca}

\author{Thaned Pruttivarasin, Michael Ramm, Hartmut H\"affner}

\affiliation{Department of Physics, University of California,  Berkeley, CA 94720, USA}

\email{hhaeffner@berkeley.edu}
\date{\today}

\begin{abstract}
We present an experimental scheme to perform spectroscopy of the $^2$S$_{1/2}-^2$P$_{1/2}$ and $^2$D$_{3/2}-^2$P$_{1/2}$ transitions in \Ca. By rapidly switching lasers between both transitions, we circumvent the complications of both dark resonances and Doppler heating. We apply this method to directly observe the micromotion modulated fluorescence spectra of both transitions and measure the dependence of the micromotion modulation index on the trap frequency. With a measurement time of 10 minutes, we can detect the center frequencies of both dipole transitions with a precision on the order of 200 kHz even in the presence of strong micromotion.

\end{abstract}

\maketitle

\section{Introduction}

Trapped ions provide an excellent experimental platform for the measurement of atomic properties such as transition frequencies, branching fractions and lifetimes of atomic energy levels. Trapped ions can be cooled easily into the millikelvin regime and are well isolated from their environment. In addition, all relevant experimental parameters such as trap frequency, magnetic and electric fields are well controlled. Because of these features, trapped ions are particularly well-suited to implement atomic clocks \cite{Rosenband} and a large body of work has been carried out to study dipole forbidden narrow atomic transitions \cite{Chwalla, Sherstov}. However, outside the context of atomic clocks, absolute frequency measurements of dipole allowed transitions are also of interest. For example, in the astrophysics community, spectra from galactic objects stem mainly from dipole allowed transitions in ions and it is desirable to compare them to precisely measured laboratory values \cite{Barbuy, Curtis}. 

\begin{figure}
\begin{center}
\includegraphics[width = 0.35\textwidth]{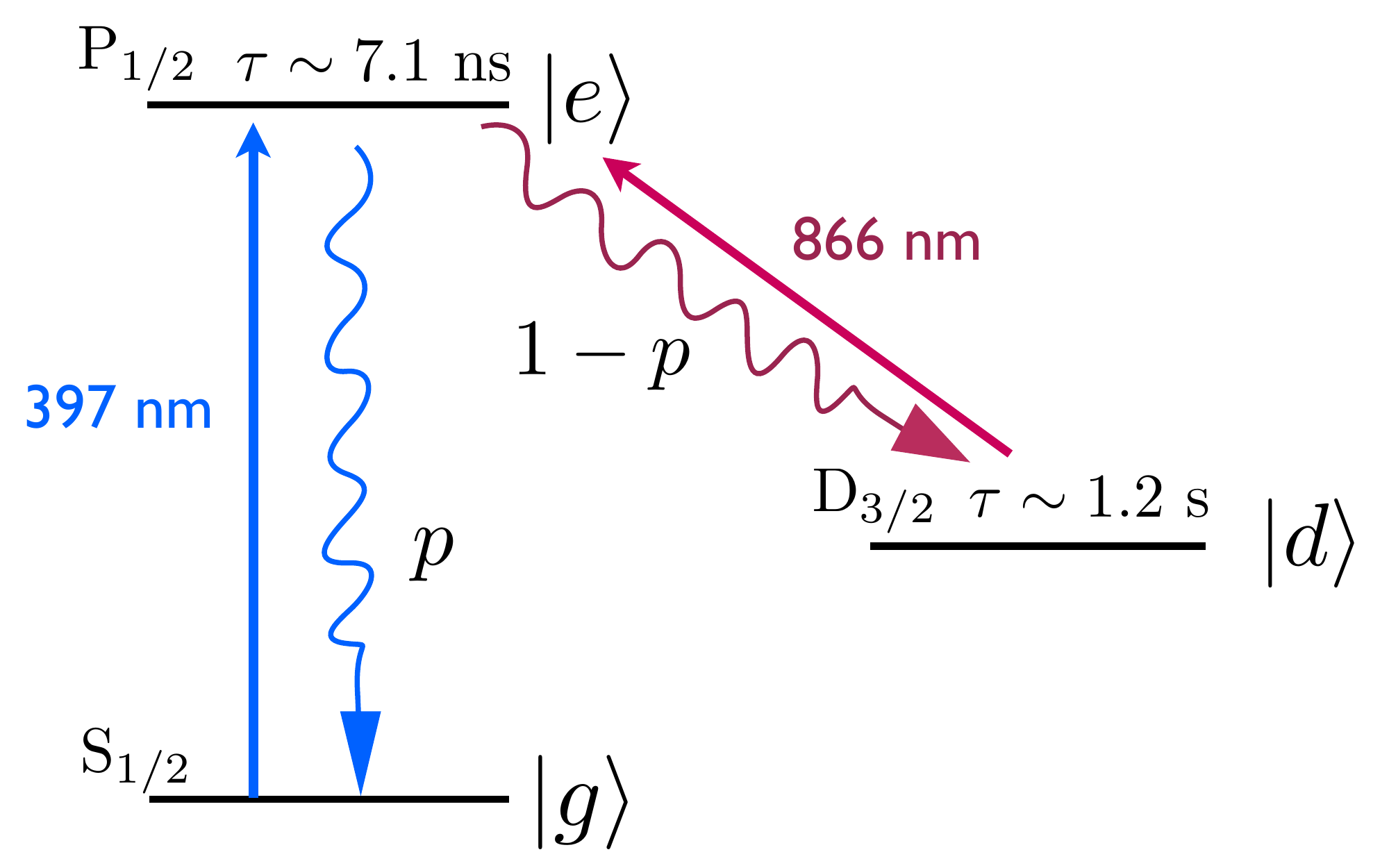}
\caption{Typical lambda-like energy level diagram as applicable to many commonly used ion species. The excited state, $\ket{e}$, can decay to either the ground state, $\ket{g}$, or the metastable state, $\ket{d}$, with probability $p$ and $1-p$, respectively. The wavelengths of the lasers that couple between states are shown for \Ca. \label{diagram}}
\end{center}
\end{figure}

A serious complication in performing spectroscopy of commonly used trapped ions is that the excited state does not only decay into the ground state but also into metastable levels (see Fig.~\ref{diagram} as an example). For instance, experiments with \Ca, \Ba, \Yb and \Sr~ions  require additional repumping lasers to depopulate those so-called dark states. However, the presence of the repumping laser strongly modifies the spectrum of the primary spectroscopy laser applied between the ground and the excited state. Moreover, changing the frequency of the spectroscopy and repumping lasers during the spectroscopy measurements modifies the cooling dynamics and influences the effective profile of the spectrum. This complication is present even in the case of two level systems. To circumvent this complication, experiments on \Mg~ions, with an essentially two-level structure, use sympathetic cooling of additional ions to maintain low temperature throughout the experiment \cite{Herrmann, Batteiger}. Another possibility is to first cool the ions and apply the spectroscopy laser only for a short duration, as carried out in \Ca \cite{Wolf}. However, for most ion species, AC-Stark effects and dark resonances due to the presence of the repumping laser still influence the obtained spectra strongly.

In this work, we present an experimental scheme that circumvents the complications of both dark resonances and Doppler heating. We apply the scheme to \Ca and perform spectroscopy of the $^2$S$_{1/2}-^2$P$_{1/2}$ and $^2$D$_{3/2}-^2$P$_{1/2}$ transitions. The scheme relies on fast photon counting and rapid switching of laser pulses similar to the one presented for the purposes of measuring the branching fractions in \cite{Ramm}. The method is applicable to ion species with a lambda-level scheme, provided that the metastable state has a significantly longer lifetime compared to the excited state. We measure the center frequency of both dipole allowed transitions with an statistical significance on the order of 200 kHz even in the presence of strong micromotion in 10 minutes of measurement time. In addition, we directly observe the micromotion modulated spectra of both the $^2$S$_{1/2}-^2$P$_{1/2}$ and $^2$D$_{3/2}-^2$P$_{1/2}$ transitions in \Ca~and measure the dependence of the micromotion modulation index on the trap frequency. 

\section{Method}

In this section, we describe the experimental scheme for performing spectroscopy on the two dipole allowed transitions presented in Fig. \ref{diagram}. In particular, we consider an excited state $\ket{e}$, which can decay back to a ground state $\ket{g}$ or to a metastable state $\ket{d}$ with probabilities of $p$ and $1-p$, respectively. 

\begin{figure}
\begin{center}
\includegraphics[width = 0.48\textwidth]{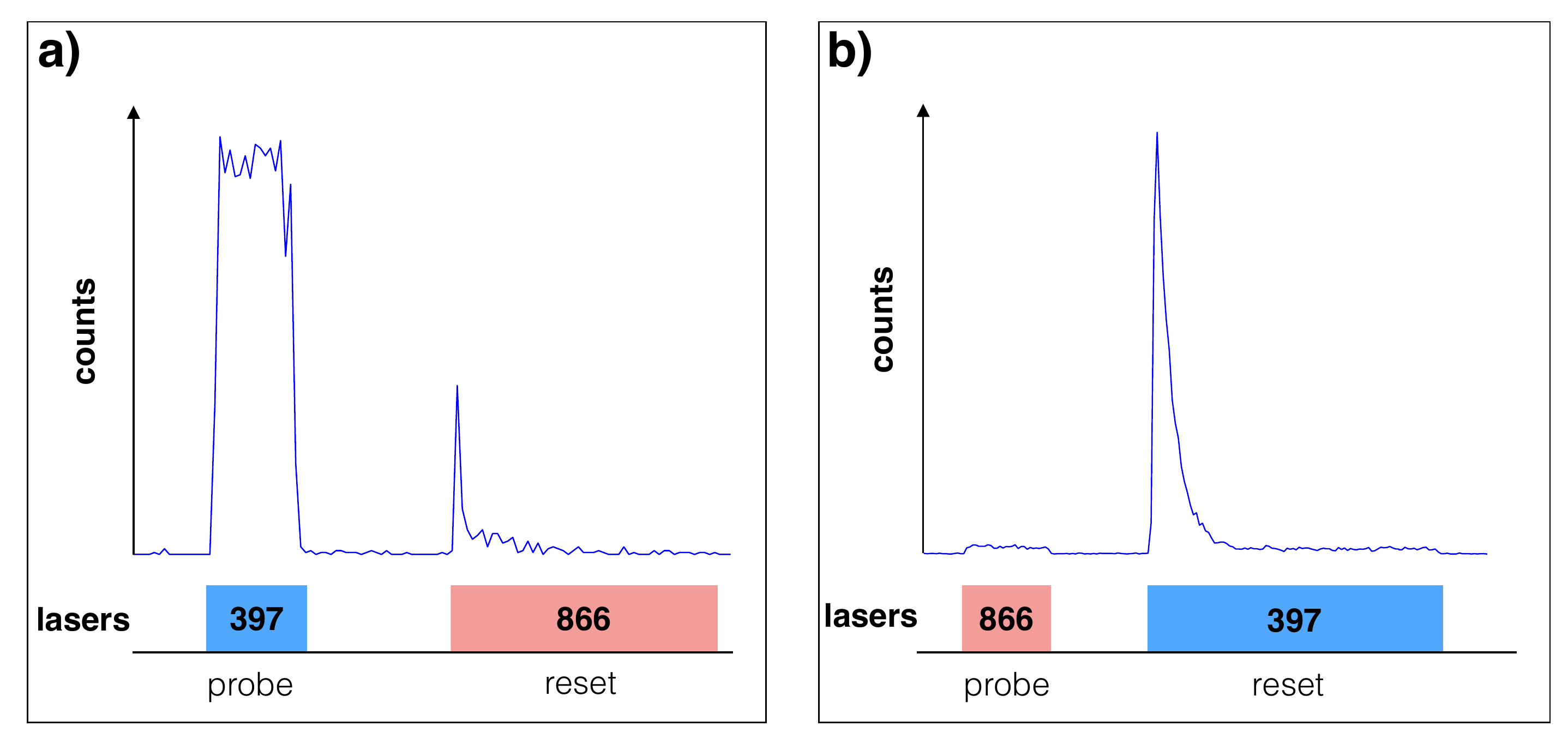}
\caption{Typical averaged fluorescence dynamics during measurement cycles. (a) Laser intensities are adjusted to perform spectroscopy on the $\ket{g}\rightarrow\ket{e}$ transition. First, the probe laser pulse at 397 nm is switched on. The following 866 nm laser pulse makes sure that all ions are in $\ket{g}$ for the next measurement cycle (reset). (b) The spectrum of the $\ket{g}\rightarrow\ket{e}$ transition is observed by collecting photons at 397 nm scattered when the 866 nm probe laser is turned on. The following 397 nm laser makes sure that all ions are in $\ket{d}$ for the next measurement. In both cases, the duration of the probe beams should be short such that the fluorescence is constant during the detection window. \label{detection_scheme}}
\end{center}
\end{figure}

First, we analyze the spectroscopy of the $\ket{g}$ to $\ket{e}$ transition. A probe laser couples these two states with the rate $R_0$, which depends on the laser detuning and intensity. Let the atom be initially in the ground state, $\ket{g}$. Then the fluorescence due to the probe laser decreases exponentially as the atom is pumped to $\ket{d}$. The experimental scheme relies on collecting the scattered photons for a short enough duration such that the scattering rate had not yet slowed down. This is achieved by making sure that the fluorescence stays constant during the detection window (see Fig. \ref{detection_scheme}). Moreover, we would like to keep the probe laser intensity low to avoid power broadening of the atomic transition. After switching off the probe light, the repumping laser is switched on to prepare the atom in $\ket{g}$ for the next probe cycle.

To describe the dynamics of the atomic populations in detail, we use a set of rate equations \cite{Cohen}:
\begin{eqnarray}
\frac{d\rho_g(t)}{dt} &=& -R_0\rho_g(t)+\left(p\gamma+R_0\right)\rho_e(t)\\
\frac{d\rho_e(t)}{dt} &=& R_0\rho_g(t)-\left(\gamma+R_0\right)\rho_e(t)\\
\frac{d\rho_d(t)}{dt} &=& (1-p)\gamma\rho_e(t),
\end{eqnarray} 
where $\gamma$ is the decay rate of the excited state and $\rho_i(t)$ is the atomic population in state $\ket{i}$. With initial conditions $\rho_g(0) = 1$ and $\rho_e(0) = \rho_d(0) = 0$, and in the limit of low saturation ($R_0/\gamma\ll1$), the time evolution of the excited state population is given by
\begin{eqnarray}
\rho_e(t) = \frac{R_0}{\gamma}e^{-(1-p)R_0t}.
\end{eqnarray}
The number of photons scattered after a duration $T$ is given by
\begin{eqnarray}
N(T) &=& \int_0^T\rho_e(t)\gamma dt \\ 
&=& \frac{1}{1-p}\left(1-e^{-(1-p)R_0T}\right).
\end{eqnarray}
We expand for $(1-p)R_0T\ll1$ and obtain
\begin{eqnarray}
N(T)&\approx& R_0T\left(1-(1-p)\frac{R_0T}{2}+...\right).
\end{eqnarray}
This shows that for $R_0/\gamma\ll1$ (low saturation) and $(1-p)R_0T\ll1$ (short probe duration), the number of photons detected in a duration $T$ is directly proportional to the coupling rate of the probe laser, $R_0$. Therefore, for a small duration $T$, we are able to perform spectroscopy of the transition with no repumping laser and without perturbations from the cooling dynamics.

Similarly, we analyze the spectroscopy of the $\ket{d}$ to $\ket{e}$ transition. A probe laser couples these two states with the rate $R'_0$. The experimental scheme is similar to the case of the spectroscopy of the $\ket{g}$ to $\ket{e}$ transition except that we do not directly detect photons scattered from the $\ket{d}$ to $\ket{e}$ transition. However, we can detect the photons emitted from $\ket{e}$ to $\ket{g}$, where the photon emission rate is directly proportional to the population in $\ket{e}$. 

We again describe the dynamics of the atomic population using rate equations:
\begin{eqnarray}
\frac{d\rho_g(t)}{dt} &=& p\gamma\rho_e(t)\\
\frac{d\rho_e(t)}{dt} &=& R'_0\rho_g(t)-\left(\gamma+R'_0\right)\rho_e(t)\\
\frac{d\rho_d(t)}{dt} &=& -R'_0\rho_d(t)+\left((1-p)\gamma+R'_0\right)\rho_e(t).
\end{eqnarray}
Since experimentally, we only detect photons emitted from $\ket{e}$ to $\ket{g}$, the number of photons emitted from this transition is exactly the population of $\ket{g}$. At low saturation limit ($R'_0/\gamma\ll1$), this is given by
\begin{eqnarray}
\rho_g(t) = 1-e^{-pR'_0t}.
\end{eqnarray}
The number of photons detected after a duration $T$ is
\begin{eqnarray}
N'(T) &=& 1-e^{-pR'_0T} \nonumber\\
&\approx& pR'_0T\left(1 - \frac{pR'_0T}{2}+...\right),
\end{eqnarray}
where we expand for $pR'_0T\ll1$. This shows that, for $R'_0/\gamma\ll1$ and $pR'_0T\ll1$, the number of photons detected from $\ket{e}$ to $\ket{g}$ transition for a duration $T$ is directly proportional to the coupling rate of the probe laser, $R'_0$.

\begin{figure}
\begin{center}
\includegraphics[width = 0.42\textwidth]{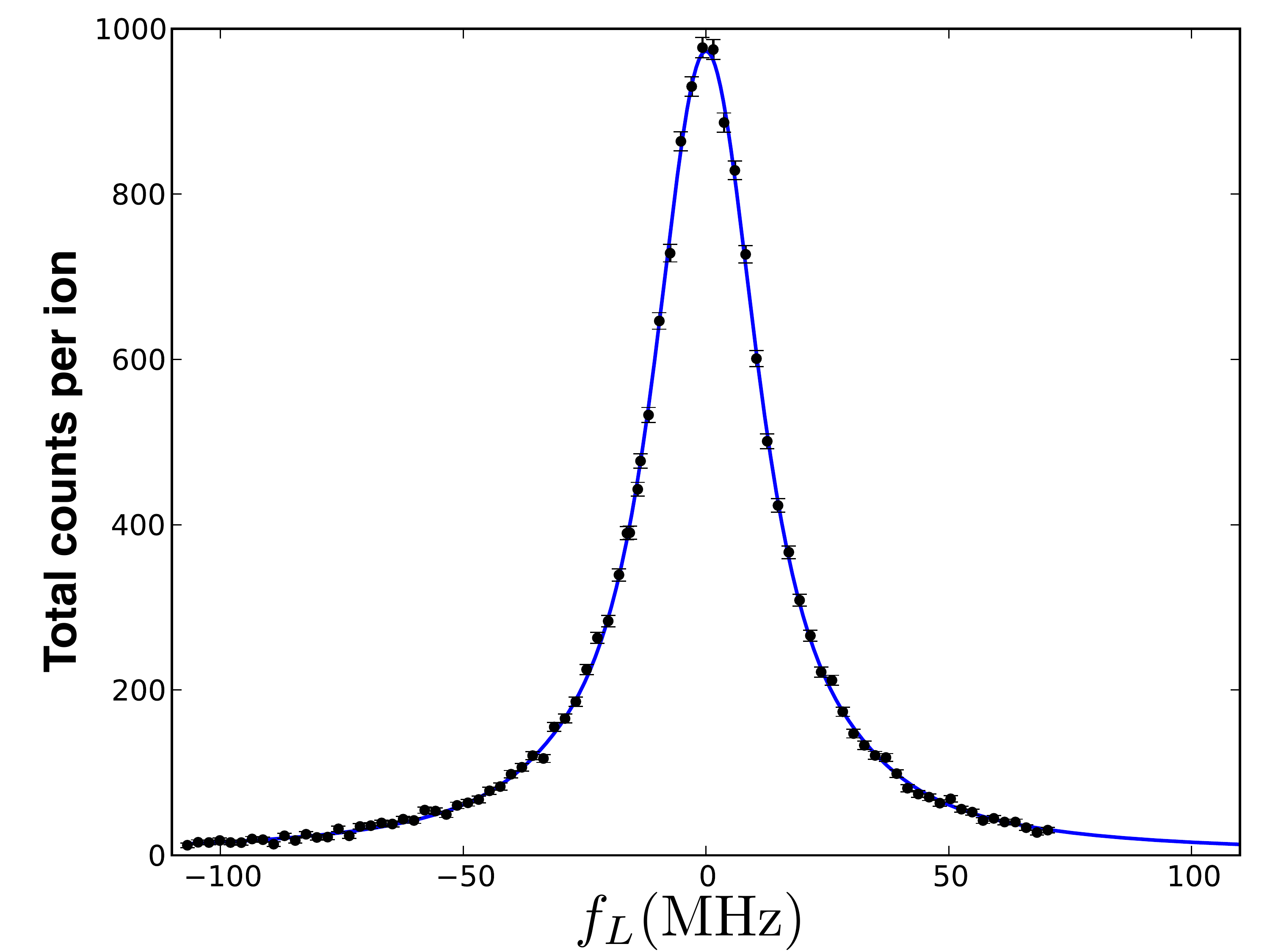}
\caption{The fluorescence spectrum of the $^2$S$_{1/2}-^2$P$_{1/2}$ transition for \Ca as the frequency of the laser, $f_\text{L}$, is scanned across the transition. After measuring for 10 minutes ($10^6$ cycles per each laser frequency), the center frequency is detected with a precision of 200 kHz. The solid line is a Lorentzian profile including the Zeeman effect and the micromotion of the ions (see text). \label{single_spectrum}}
\end{center}
\end{figure}

We apply this method to measure fluorescence spectra of trapped \Ca. In this case, the laser wavelengths that couple $\ket{g}\rightarrow\ket{e}$ and $\ket{d}\rightarrow\ket{e}$ are at 397 nm and 866 nm, respectively  (see Ref.~\cite{Ramm} for details on the experimental apparatus). To observe the fluorescence spectrum for the $\ket{g}\rightarrow\ket{e}$ transition, in each measurement iteration, the 397 nm laser (the probe) is first turned on for a short duration. We adjust the intensity and the duration of the probe beam such that the fluorescence from the ions is constant during the probing time, as shown in Fig. \ref{detection_scheme}a. This ensures that the probability that ions are excited is small. After switching off the probe light, the ions are reset to $\ket{g}$ with a strong pulse at 866~nm, preparing all the ions at the ground state for the next iteration. 

To observe the fluorescence spectrum for the $\ket{d}\rightarrow\ket{e}$ transition, the roles of the 397 nm and 866 nm are reversed, as shown in in Fig. \ref{detection_scheme}b. In both cases, only the 397 nm photons scattered from ions are collected during the probe duration. For our measurement, the probe beam duration is typically 2 $\mu\text{s}$ and the reset beam duration is typically 10 $\mu\text{s}$. For switching the laser intensities, we use acousto-optical modulators (AOMs) in a double-pass configuration.

To minimize heating and cooling effects of the ions from due to the probe and reset laser, we perform Doppler cooling a period of 1 ms after every 50 repetitions of the measurement cycle.  Since the fluorescence is constant during the detection window, on average, we scatter less than one photon from the probe laser during each measurement cycle, the ions remain close to the Doppler temperature throughout the experiment.

Fig. \ref{single_spectrum} shows the spectrum obtained with this method on the $^2$S$_{1/2}-^2$P$_{1/2}$ transition.  The method is not restricted to a single ion, which allows us to increase signal-to-noise by performing measurement with many ions. For the data shown in Fig. \ref{single_spectrum}, we use 7 ions and $\sim$$10^6$ cycles per each probe laser frequency. This corresponds to a total of measurement time of 10 minutes. The solid line is a Lorentzian profile with a natural linewidth of 22.4 MHz \cite{Jin}. We also include line broadenings from the Zeeman effect due to an applied magnetic field of $\sim$0.8 gauss and the micromotion, which will be discussed in details in the next section. The fit yields $\chi^2_\text{red} \approx 1.05$, and the center of the spectrum is detected with a precision on the order of $\sim$$200$ kHz.

\section{micromotion modulated spectrum of trapped ions}

In this section we study micromotion modulated spectra of trapped ions of both $^2$S$_{1/2}-^2$P$_{1/2}$ and $^2$D$_{3/2}-^2$P$_{1/2}$ transitions in \Ca~using the method described in the previous section. 

To describe the effect of the micromotion caused by a radio-frequency (RF) drive in a linear Paul trap, we follow the analysis by Berkeland \textit{et al.} \cite{Berkeland}. We consider a case where the micromotion of the ions is dominated by the excess micromotion in the radial directions caused by a stray static electric field. We also include a possibility of having an additional micromotion caused by a relative phase difference of the potentials between trap RF-electrodes.

Due to the Doppler effect, the electric field of the probe laser as seen by ions undergoing micromotion is given by
\begin{eqnarray}
E(t) = \text{Re}\left[E_0 e^{i(\vec{k}\cdot(\vec{u}_0+\vec{u}')-\omega_\text{laser}t)}\right],\label{eq_laser_field}
\end{eqnarray}
where $E_0$ is the amplitude of the laser field, $\vec{k}$ is the wavevector of the laser, $\omega_\text{laser}$ is the frequency of the laser, $\vec{u}_0$ and $\vec{u}'$ are the displacements of the regular motion and excess micromotion, respectively. At small excess micromotion amplitude, we can write
\begin{eqnarray}
\vec{k}\cdot\vec{u}' = \beta \cos{(\Omega t +\delta)},\label{eq_amplitude}
\end{eqnarray}
where $\Omega$ is the trap drive frequency, $\beta$ is the modulation index of the micromotion and $\delta$ is the associated phase. The dependence of the micromotion modulation index on the two radial trap frequencies, $f_x$ and $f_y$, is given by Eq. (22) in \cite{Berkeland}:
\begin{eqnarray}
\beta = \sqrt{\frac{1}{\lambda_L^2\Omega^4}\left(\frac{A}{f_x^2}+\frac{B}{f_y^2}\right)^2+C},\label{eq_beta_trap}
\end{eqnarray}
where $\lambda_L$ is the wavelength of the probe laser, $A$ and $B$ are constants proportional to the projection of the stray electric field on the two radial directions and $C$ describes additional micromotion from a relative phase difference of the potentials between trap electrodes. From Eq. (\ref{eq_laser_field}) and (\ref{eq_amplitude}), the fluorescence of the ion at low probe laser intensity can be expanded using Bessel functions:
\begin{eqnarray}
P \propto \|{E_0}\|^2 \sum_{n=-\infty}^{\infty}\frac{J_n^2(\beta)}{(\Delta + n\Omega)^2+(\gamma/2)^2},\label{eq_lineshape}
\end{eqnarray}
where $J_n$ is the Bessel function of $n$-th order, $\Delta$ is the detuning of the laser frequency from the atomic transition and $\gamma$ is the decay rate of the excited state. We also account for additional line splittings due to the Zeeman effect by summing over all possible magnetic sublevels.

\begin{figure}
\begin{center}
\includegraphics[width = 0.5\textwidth]{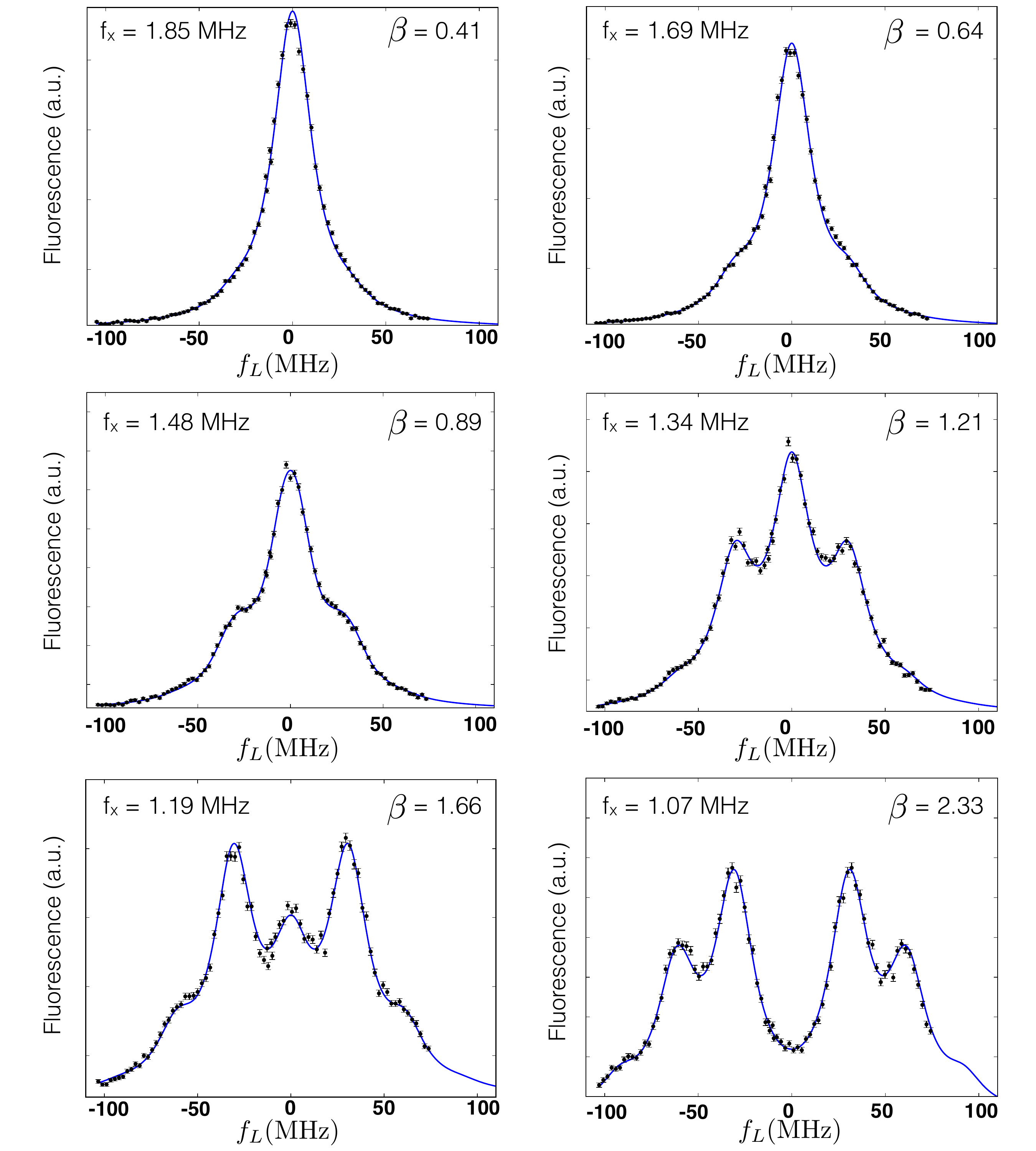}
\caption{Micromotion modulated fluorescence spectra of the S$_{1/2}\rightarrow$P$_{1/2}$ transition for \Ca~at different modulation indices, $\beta$, due to different trap frequencies.\label{blue_spectrum}}
\end{center}
\end{figure}

\begin{figure}
\begin{center}
\includegraphics[width = 0.35\textwidth]{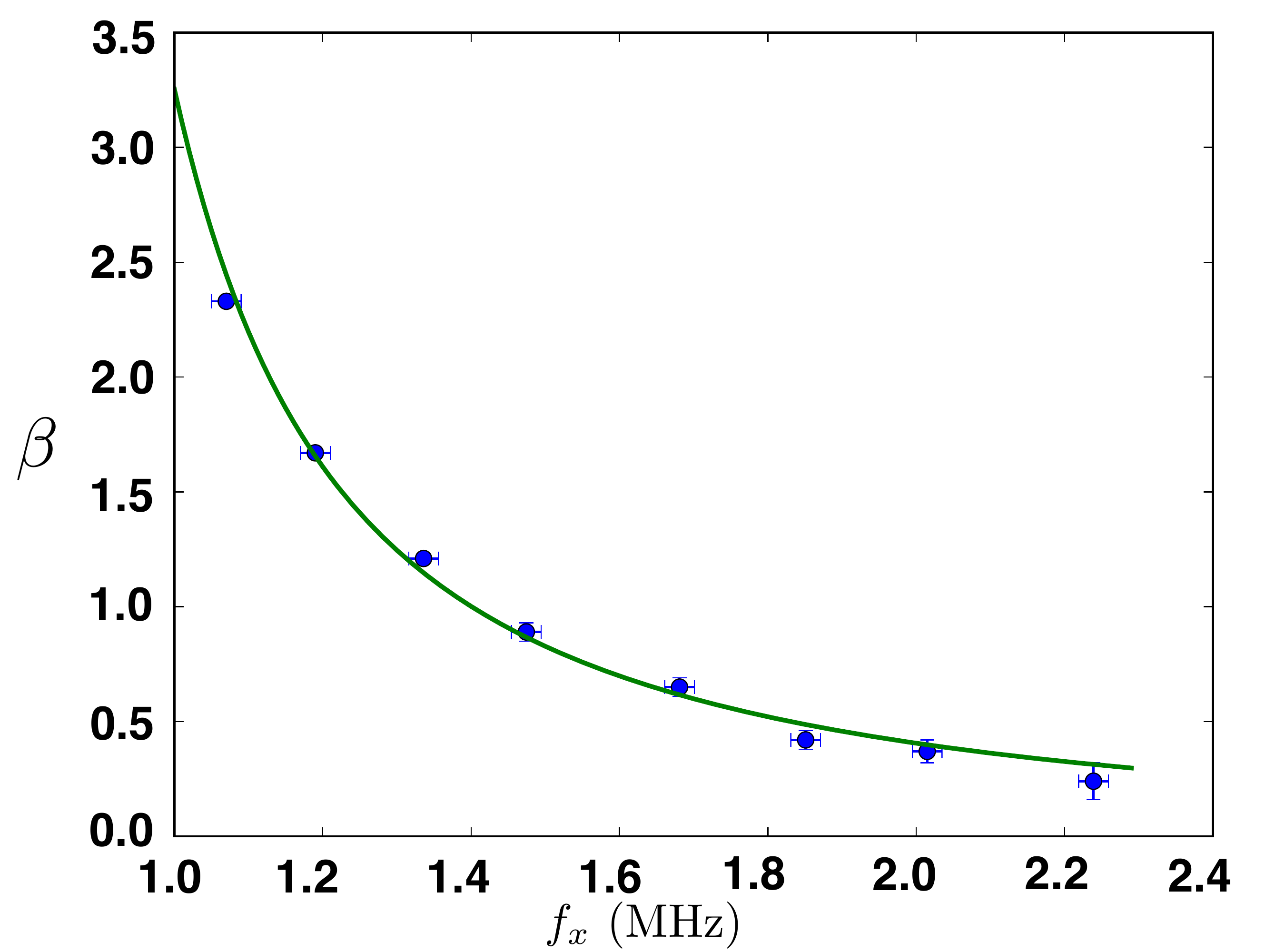}
\caption{Micromotion modulation index, $\beta$, measured as a function of trap frequency, $f_x$. The line is the theoretical fit according to Eq. (\ref{eq_beta_trap}) with the value of the radial trap frequencies given by Eq. (\ref{eq_trap_dependence}). The errorbars in $\beta$ are comparable to the size of the markers. \label{beta}}
\end{center}
\end{figure}

We use the experimental scheme described in the previous section to observe the modulations of fluorescence spectra of the $^2$S$_{1/2}-^2$P$_{1/2}$ transition from ion micromotion.  We measure the modulation index, $\beta$, as a function of the radial trap frequencies. By changing the amplitude of the RF-drive applied to the trap electrodes, the relationship between the two radial trap frequencies is measured to be
\begin{eqnarray}
f_y  = 1.18(4) \cdot f_x  - 0.63(8) \text{     (MHz)}. \label{eq_trap_dependence}
\end{eqnarray}
The fluorescence spectra obtained for different trap frequencies, $f_x$, fitted to the model given in Eq.~(\ref{eq_lineshape}), are shown in Fig. \ref{blue_spectrum}. The micromotion sidebands at the drive frequency ($\Omega$ $\sim$ $2\pi\times 30.7 \text{ MHz}$) are clearly visible. The dependence of the modulation index, $\beta$, on $f_x$ is plotted in Fig. \ref{beta}. The data is fitted to a the model according to Eq. (\ref{eq_beta_trap}) and (\ref{eq_trap_dependence}) with the errorbars in $\beta$ (comparable to the size of the markers) derived from the fit in Fig. \ref{blue_spectrum}. The fit yields $(A,B) =(2.6\pm1.5,2.6\pm0.4)\times 10^{5} \text{ nm}\cdot\text{MHz}^4$ and $C=0.00(3)$ with $\chi^2_\text{red} \approx 1.06$. The value of $C$ suggests that the effect from the phase difference of the potentials on the trap electrodes is small.

\begin{figure}
\begin{center}
\includegraphics[width = 0.5\textwidth]{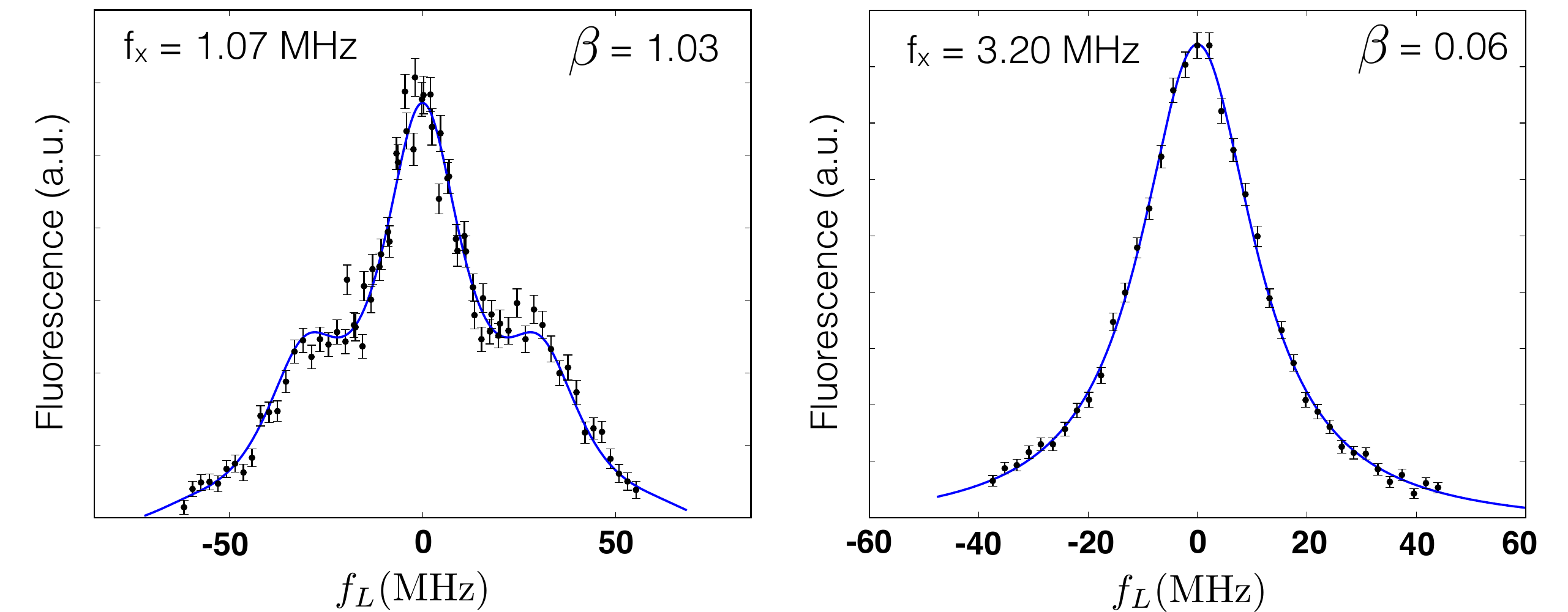}
\caption{Micromotion modulated fluorescence spectra of the D$_{3/2}\rightarrow$P$_{1/2}$ transition for \Ca at high and low modulation index $\beta$ due to different trap frequencies. \label{red_spectrum}}
\end{center}
\end{figure}

We also observe the micromotion modulated fluorescence spectra of the $^2$D$_{3/2}-^2$P$_{1/2}$ transition, as shown in Fig. \ref{red_spectrum}, which are measured at two different trap frequencies. At the trap frequency of $f_x = 1.07 \text{ MHz}$, the spectra of the $^2$D$_{3/2}-^2$P$_{1/2}$ and $^2$S$_{1/2}-^2$P$_{1/2}$ transitions yield the modulation indices to be $\beta_{866} = 1.03(5)$ and $\beta_{397}=2.33(5)$. The ratio $\beta_{397}/\beta_{866} = 2.26(12)$ is consistent with the ratio of the laser wavelengths $\lambda_{866}/\lambda_{397} \approx 2.18$ according to Eq. (\ref{eq_beta_trap}).

Since we are able to directly observe the effect of micromotion on the fluorescence spectra, this method can be used to compensate excess micromotion in the direction along the laser beam direction, $\vec{k}$. The projection of the micromotion can be on either the main Doppler cooling laser (397 nm for \Ca) or the repumper laser (866 nm for \Ca).

\section{Conclusions}

In this work, we present an experimental scheme to perform direct spectroscopy on the $^2$S$_{1/2}-^2$P$_{1/2}$ and $^2$D$_{3/2}-^2$P$_{1/2}$ transitions for \Ca. The method allows us to circumvent the usual limitations due to a lambda three-level energy structure present in many species of trapped ions. We apply this method to directly observe micromotion modulated spectra of the fluorescence for both transitions in \Ca~and show that the dependence of the micromotion modulation index on the trap frequency agrees well with the model presented in \cite{Berkeland}. Even in the presence of strong micromotion, we detect the center frequency with precision of the order of 200 kHz, which is comparable to or better than recent measurements on dipole transitions of trapped ions \cite{Wolf, Wolf2, Herrmann,Wan, Batteiger}. The absolute transition frequencies of both transitions could be determined with a frequency comb to within the same order of precision \cite{Wolf, Herrmann}. Specifically, the absolute transition frequency for the $^2$D$_{3/2}-^2$P$_{1/2}$ of \Ca~(866 nm) has not yet been determined to this level of precision \cite{Hempel}. The method is easily applicable to other species of trapped ions and the number of lasers required is minimal.

\section{Acknowledgement}

This work was supported by the NSF CAREER program grant \# PHY 0955650. MR was supported by an award from the Department of Energy Office of Science Graduate Fellowship Program (DOE SCGF). 

\appendix

\end{document}